# Quantized amplitudes in a nonlinear resonant electrical circuit

B. Cretin, D. Vernier
FEMTO-ST, department LPMO - UMR6174
Université de Franche-Comté, CNRS, ENSMM, UTBM
32, avenue de l'Observatoire, 25044 Besançon cedex France

**Abstract**
We present a simple nonlinear resonant analog circuit which demonstrates quantization of resonating amplitudes, for a given excitation level. The system is a simple RLC resonator where C is an active capacitor whose value is related to the current in the circuit. This variation is energetically equivalent to a variation of the potential energy and the circuit acts as a pendulum in the gravitational field. The excitation voltage, synchronously switched at the current frequency, enables electrical supply and keeping the oscillation of the system. The excitation frequency has been set to high harmonic of the fundamental oscillation so that anisochronicity can keep constant the amplitude of the circuit voltage and current. The behavior of the circuit is unusual: different stable amplitudes have been measured depending on initial conditions and excitation frequency, for the same amplitude of the excitation. The excitation frequency is naturally divided by the circuit and the ratio is kept constant without external disturbance. Moreover, a variation of the dumping does not affect significantly the amplitudes as long as the oscillation is observed. And lastly, electrical pulses can change, as in quantum systems, the operating amplitude which is auto-stable without disturbances. Many applications of this circuit can be imagined in microelectronics (including computing), energy conversion and time and frequency domains.

**Introduction**
Almost all real systems are nonlinear and it is well known that nonlinearity needs complex analysis. Dynamics of nonlinear system can yield to chaotic behavior but, depending on the system and on its excitation, stable cycles can be observed. These stable cycles can have different energies and the levels are quantized on the macroscale. Historically, the first studied nonlinear systems where based on mechanics. The basic example taken for nonlinear oscillator has been the pendulum which served as time reference at low oscillation level, *i.e.* in approximate linear regime, for about three centuries [1]. Obviously, the oscillation of any non linear system can be expressed as a sum of harmonic (Fourier series) oscillations with frequencies which are integer multiples of the fundamental frequency. Two main effects are encountered in non linear oscillating system: the foldover effect which has got its name from the bending of the resonance curve peak in the amplitude versus frequency plot and, more interesting for the proposed topic, superharmonic resonance is simply the resonance with one of this higher harmonics of a nonlinear oscillation. The behaviors of the nonlinear systems have extensively been studied [2-8] on both theoretical and experimental levels. Specific studies of time and frequency applications have shown original results [9-10]. In this work we study the effect of specific excitation with a gated signal. We show that additional resonance peaks can be expected. Another approach should also be considered: in non linear systems, chaos behavior is often encountered, but in the case of high Q-factor and sine excitation, synchronous oscillations give stable amplitudes which are sensitive dependent on initial conditions. These stable amplitudes are related to the presence of stable circles in the phase space.

In this paper, we describe the association of nonlinear resonant electrical circuit and synchronous excitation which enables different stable amplitude relatively independent on the excitation voltage. In nonlinear circuits, due to huge foldover effect (non bijective amplitude versus frequency curve yields to bistability and hysteresis), two stable amplitudes can be obtained for the same excitation frequency which is close to the fundamental frequency. In our setup the excitation is at odd harmonics of the natural oscillation frequency and the signal is gated, which is unconventional. The actual circuit is formally the equivalent of the Doubochinsky's pendulum [11] which has been excited with an electromagnetic coil, but the use of high frequency and simply adjustable parameters gave quickly various oscillation configurations and a good overview of such a system (with a pendulum having a Q-factor about 200 and 1 second period, more than 10 minutes are needed to obtain a good stabilization of the oscillation; waiting 5Q periods enable 1% accuracy). In order to avoid digital quantization effects, we have designed a completely analog circuit which can be easily understandable and modeled. This circuit is a simple RLC resonator where C is an active capacitor whose value is related to the current in the circuit. A gate shaped driving has been included in order to synchronously excite the oscillation. Because, an exhaustive study is interesting, some potentiometers have been inserted to allow the preset of the values of the main parameters (gain, gate width, non linearity). The frequency has been chosen about 1 kHz for real time response. Obtained results have demonstrated quantized amplitudes, even at first odd harmonics of the natural frequency of the oscillation, depending on the initial conditions provided with a low frequency generator. We report some results demonstrating the interest of the concept for self-stabilized devices on the macro- or microscale. Some potential applications are given.

**1. Physical basis**
It is well known that no model of a real system is truly linear. Most phenomena are profitably studied as linear approximations to the real models, mainly because nonlinear models need a very good knowledge of mathematics, and in the case of oscillating systems, of the nonlinear differential equations which cannot be solved analytically. Digital computing yields solutions of these differential equations but the physicist is not able to evaluate quickly the influence of the different parameters. This limitation has been a brake for the development of nonlinear physics. Moreover, many nonlinear behaviors are not completely known and only approximately modeled. Lastly, for numerous applications, nonlinearity is avoided: this is the case for springs in mechanics, for oscillators in clocks (in this case the nonlinearity is fundamental but it is usually low enough), for current and voltages in electrical circuits,…
We have based this study on nonlinear oscillating circuits. The most well-know nonlinear basic equation is the one of the pendulum:

$$\frac{d^2\theta}{dt^2} + \beta\frac{d\theta}{dt} + \omega_0^2 \sin\theta = 0, \tag{1}$$

where θ is the instantaneous angle of the pendulum, β the damping (usually from air drag), and t the time.
For low values of β, the period of the motion can be approximated by using elliptic integral of first kind K(x) [12]. The relative period is expressed as:

$$\frac{T}{T_0} \approx \frac{2}{\pi} K\left[\sin\left(\frac{\theta_0}{2}\right)\right], \tag{2}$$

where $T_0$ is the period of the pendulum swinging through a small angle elongation $\theta_0$; the Jacobian elliptical function K(x) has been tabulated [13]. For instance, for 90° angle, we obtain $T/T_0 \approx 1.18$. The period of the pendulum clearly depends on the amplitude and this anisochronous behavior enables to stabilize the amplitude of the pendulum excited with a sine force. The same non linear behavior has been encountered in other field of the physics and on different scale: for instance non linear oscillators are found in scanning probe microscopes (SPM, [14]). We have used this feature in our system, but we had a specific excitation in order to synchronize the system and to keep constant the swinging elongation.
Concretely, our serial RLC circuit can be modeled by the following equation:

$$L\frac{\partial^2 i}{\partial t^2} + R_s\frac{\partial i}{\partial t} + \frac{i}{C} = A.g(i)\sin(\omega t), \tag{3}$$

where *i* represents the electrical current, *L* the inductance, *R$_S$* the serial resistance of the circuit (representing the losses), *C$_0$* the capacitance, *A* a constant, *ω* the angular frequency of the excitation and *g(i)* a gate function so that:

$$g(i) = 1 \text{ if } |i| \leq |i_0| \text{ and } g(i) = 0 \text{ if } |i| > |i_0| \qquad (4)$$

This is obviously the equation of a standard RLC circuit. The only difference is the gated excitation. In order to approach the behavior of the pendulum, we have used analog computing circuit to obtain the appropriate value of the capacitance:

$$C = C_0 \frac{i}{\sin i}, \qquad (5)$$

where $C_0$ is constant. The equation (4) is expressed:

$$\frac{\partial^2 i}{\partial t^2} + \frac{R_S}{L}\frac{\partial i}{\partial t} + \omega_0^2 \sin(i) = \frac{A}{L} g(i)\sin(\omega t) \qquad (6)$$

$$\text{with } LC_0\omega_0^2 = 1$$

The solution of this equation cannot be completely obtained by analytic calculations because the excitation term of the equation is a gated signal at angular frequency ω which is usually a high harmonic of the natural angular frequency ω$_0$ of the RLC circuit oscillating at low amplitude.

An approximate value of the amplitudes a$_n$ of oscillation has been given in [11] by using averaging method:

$$a_n \approx B\sqrt{8\left(1-\left(\frac{\omega}{n\omega_0}\right)^2\right)}; \text{ with } n = 2m+1, m = 0,1,2,..., \qquad (7)$$

where B is a constant. A phenomenological interpretation can be easily given. If we consider the frequency response of nonlinear systems, the possible shapes are shown in Fig. 1. If the system is linear or quasi linear, the conventional resonance curve is obtained. The curve 3 corresponds to unconventional electrical or mechanical systems and will not be discussed here (experimentally, our electronic circuit enables this behavior but results are not reported here). For most non linear conventional oscillators, the potential well has a cosine shape (for the pendulum or our circuit, the potential energy is expressed: E(x)=1-cos(x), x being the variable; the first derivative of E(x) is sin(x), found in the equation) and the curve is folded on the left side (curve 2). In the frequency range between the two arrows, the hysteretic and bistability effects are obvious; the foldover effect and superharmonic resonance result from the shape of this curve. Experimentally, this means that two different amplitudes can be obtained for the same frequency. Furthermore, as the amplitude of oscillation gets larger the period gets longer (frequency decreases); this is a fundamental effect of nonlinearity in curve 2. We have to notice that this behavior is different from the result given by Eq. (7), where the excitation frequency is not the oscillation frequency. Equation (7) shows many possible stable amplitudes corresponding to different oscillation frequencies for the same excitation frequency which is an odd harmonic of f$_0$, the basic frequency of oscillation of the system. The odd numbers are explained by the sign of the excitation current which depends on the sign of the current in the RLC circuit, *i.e.* on each half-cycle.

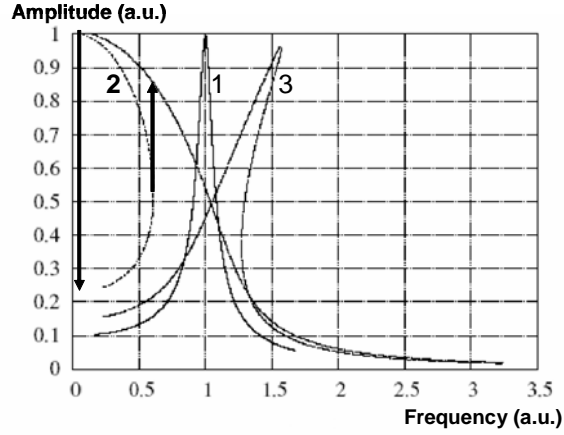

Figure 1. Resonance shape of oscillating system depending of the nonlinearity

## 2. Principle of the nonlinear RLC circuit

In order to correctly design this analog circuit, we have first given a functional scheme separating the different blocks. A diagram of the system is drawn in Fig.2. The inductance is capital for the circuit because we need a high Q factor (typically above 100 at 1 kHz, order of magnitude of frequency $f_0$). We have built the inductance using a ferrite core. The coil has been designed to obtain a value of 100 mH. Experimental Q factor is about 160 at 1 kHz (measured with RLC meter Philips PM6303). In order to sense the current, a small (1 Ω) resistance R has been inserted in the circuit. A differential amplifier directly gives the appropriate voltage proportional to the current in the RLC circuit. A comparator drives the switch which enables the opening of the input for the low frequency generator. The value of the reference current $i_0$ can be preset with a potentiometer. The switch is opened only when $|i| \leq |i_0|$. In this case (opening time small compared to the period), the small resistance of the generator is not a source of disturbance for the oscillation. A non linear circuit generates the function of Eq. (5). It is based of conventional wave shaper circuits using resistors and diodes (Piecewise linear approximation).

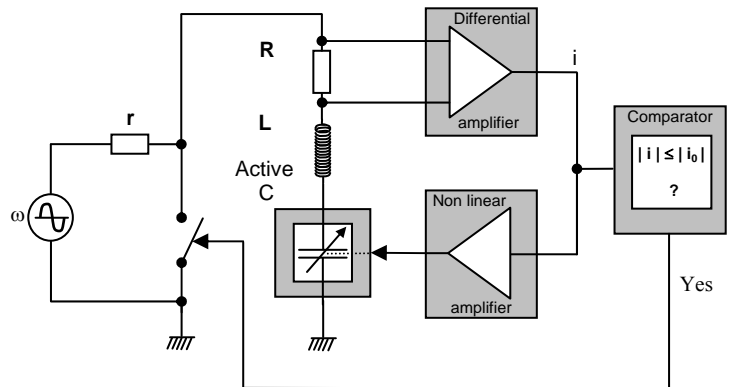

Figure 2. Functional diagram of the non linear circuit

The voltage controlled capacitance is based on the principle given in Figure 3. Two operational amplifiers are used in a gyrator configuration that simulates a capacitance whose value is expressed as:

$$Céq = \frac{(R_{30} + R_{29})}{R_{30}} \cdot \frac{R_{27}}{R_{25}} \cdot C_{24} \qquad (8)$$

Resistor $R_{27}$ is the main component for voltage control. In order to control the capacitance, a transconductance amplifier LM13700 has been used to simulate a floating resistance driven by the voltage coming from the nonlinear amplifier. Simplification of Eq. 8 is easily obtained by choosing $\boldsymbol{R_{29} = R_{30}}$. The experimental value of $R_{27}$ is in the 14kΩ - 25kΩ range, that is enough to vary the period in 1.3 ratio; this obviously limits the i/sin(i) curve but is not really restrictive (to give an equivalence, this variation of the period corresponds to 110° amplitude $θ_0$ for a simple pendulum). One have to point out that the simulated capacitance is not perfect and the equivalent resistance is a cause of Q-factor decrease; electronic compensation of these losses are source of instabilities and we have preferred to keep good oscillation conditions with a lower Q-factor.

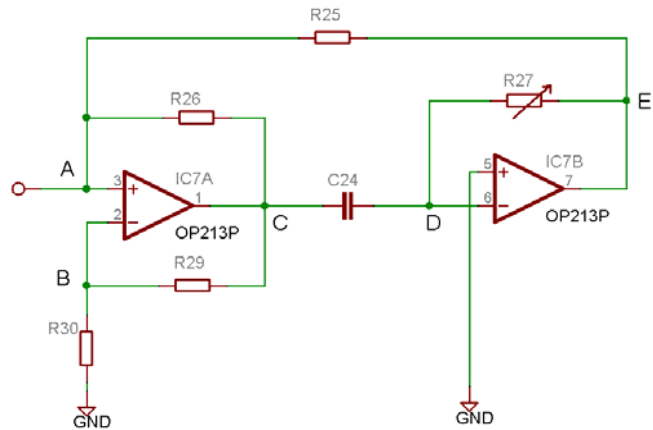

Figure 4. Diagram of the simulated capacitance (between point A and GND) used in the nonlinear circuit

## 3. Experimental results

The circuit has been supplied and tested in different configurations. We have first checked the non linear behavior of the resonance. In order to easily drive the RLC resonator, we have added a second coil to the inductance so that it acts as a simple transformer and enables the preset of the oscillations. With the chosen coil (L=100 mH) and capacitance (measured value about 0.38μF for low signal levels), the resonant frequency has been found to be 819Hz, i.e. close to the expected value. The standard excitation has been shunted in order to assure the operation of the RLC circuit only, without closed loop effect. The bending of the resonance curve peak in the amplitude versus frequency plot is clearly visible in Fig. 5 where the response is plotted for different levels of the excitation (the excitation has been done through the secondary coil of the inductance; a 1 kΩ serial resistance has been inserted to avoid significant variation of the Q-factor of the circuit. At these levels, and for the measured Q-factor (about 15, depending on the amplitude), the foldover effect is not a source of bistability as shown in the figure (for the highest amplitudes, the curves have been plotted up and down to check the hysteretic effect and bistability). Consequently, it is obvious that any bistability of the circuit originates in another effect (experimentally the -15/+15V supply limits the maximum value to 13V approximately, i.e. the maximum amplitude plotted). As expected, the nonlinearity decreases the resonant frequencies as the amplitude increases.

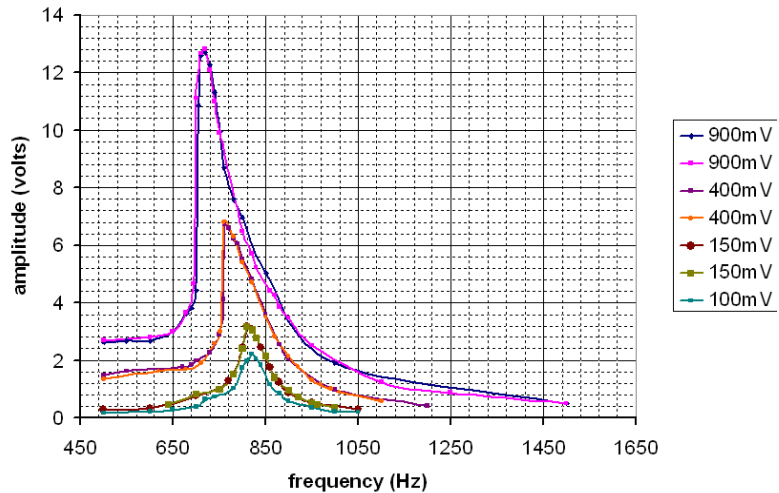

Figure 5. Frequency response of RLC circuit without switching for different values of excitation voltages

In a second step, we have kept the same generator (AGILENT Model 33120A) and used another one (same model which synchronization input) to understand the operation of the closed-loop circuit. We have used a second generator for the excitation, at a frequency that is an odd harmonic of the oscillation. This configuration has allowed us to preset the different potentiometers and to well understand the operation of the circuit. The first experiments have demonstrated the fundamental effect of the gate width that should be correctly preset in order to permit the immediate oscillation of the system when the initial conditions are correctly chosen. An oscilloscope (model AGILENT Infinium DSO 80204B) has been used to record the different signals that are interesting for the understanding of the operation. At low excitation level, we have recorded both gated excitation signal (voltage switch) and capacitance voltage). Figure 6 shows an example of oscillogram that has been obtained in the case of excitation with two generators (one for the RLC circuit directly on the secondary coil of the inductance and the second one on the switch input) to determine the optimized preset values. Amplitude values are not given because the figure serves as an illustration of the process (the circuit is not "self-oscillating" and the higher the excitation, the higher the amplitude is). In this case, the high frequency is exactly 7 times the resonant frequency of the RLC circuit. Before each experiment, we have checked the correct preset values of the parameters by using this method. The advantage of the electrical circuit compared to a mechanical one is obvious: all the signals are easily monitored and the frequency provides real time response.

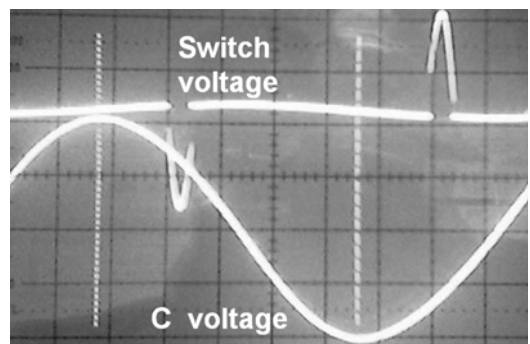

Figure 6. Oscillogram showing the gated excitation voltage and
the capacitance voltage in the case of excitation with two generators

These first experiments also led us to add a LF power amplifier to assure the oscillation of the circuit when excited through the analog switch. After the necessary preset, we have driven the circuit in the expected excitation mode (odd harmonic of the "natural" frequency of the RLC circuit; in practice, we have checked it for different harmonics above $5f_0$). The first start gave no oscillation because the switch was kept opened and the averaged voltage was null. This was coherent with the theory, because the circuit needs specific initial conditions to oscillate. After that, we have connected the second generator to the secondary coil of the inductance. Depending on the amplitude level, when we stop it, different oscillation levels can be observed. A long analysis of the oscillogram has shown that the oscillation is kept perfectly stable when no external perturbation is applied. We never observed any chaotic behavior of the system. If the oscillation is stopped, the amplitude is kept null and a new pulse is necessary to restart the oscillation. Moreover, we have verified that the excitation amplitude weakly affects the oscillation as long as it is high enough to compensate the energy losses (equivalent loss resistor of the RLC circuit). With our low Q-factor (about 15), typical variations of the amplitude are a few percents when the excitation amplitude is doubled; this result is clearly explained by the variation of the oscillation phase lag that tends to keep the same value of the averaged gated signal.

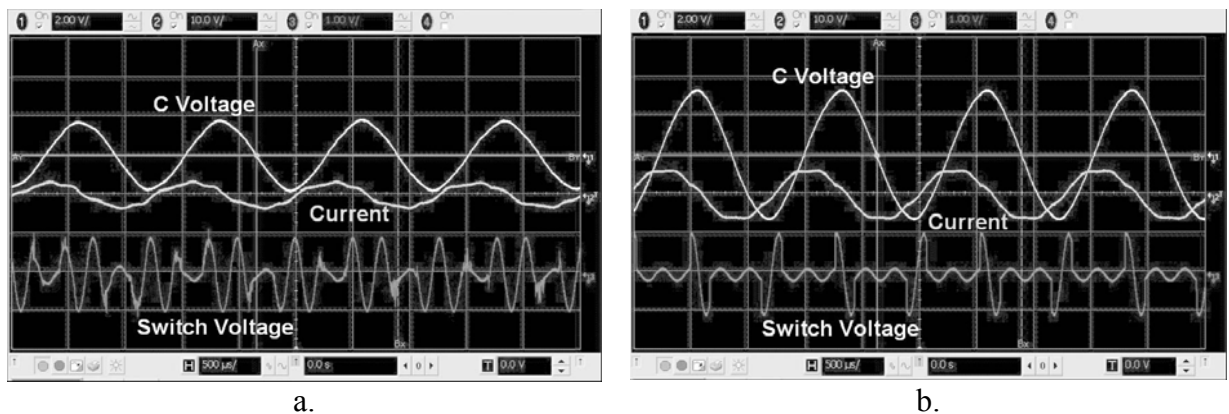

Figure 7. Different signals obtained for the same excitation voltage and frequency, depending on initial conditions

The more significant result is the memory effect related to the multi-stability effect. It is demonstrated in Figure 7 where the excitation voltage and frequency is kept constant. The frequency has been chosen a little bit below $5f_0$ because the frequency decreases at high amplitude levels: exactly 3940 Hz). Depending on the initial conditions, the circuit is oscillating with the amplitude observed in Fig.7.a or 7.b (or null if no pulse is applied to start the oscillation). One may notice that the number of half cycles needed for the excitation is odd, as expected, and is different in the two configurations. The gate function is exactly the same (the opening of the switch is related to the value of the current: the lowest the current, the highest the opening time of the gate function. This result is significant of the non linear behavior of the system: first, obviously, the two oscillation frequencies are different, in accordance with the theory (respectively 805 and 788 Hz, the last value corresponding to the highest amplitude as shown in Figure 5). Secondly, the oscillation is kept on the same level, even if the voltage of the excitation is reasonably varied. These features are the base of new applications of RLC circuits: multi-state memories, down-frequency conversion, high efficiency ac to ac (or dc by using rectifiers) converters,…

**Conclusion and prospects**

We have demonstrated that multiple stable amplitudes can be easily obtained with a simple nonlinear RLC circuit driven through a synchronous switch at frequencies harmonic of the natural frequency of the resonator. Different interesting features have been demonstrated for a relatively low Q-factor (about 15) and below 1 kHz resonant frequency:

a. the circuit should be driven with odd harmonics so that the average current of each pulse of the gated signal is not null and of the sign of the current.

b. the circuit is not self-oscillating when driven with a sine source and the amplitude depends on the initial conditions (energy in the circuit at starting time) for a given amplitude and frequency of excitation. Different amplitudes are observed and the corresponding frequency can be explained by anisochronicity and not by the simple bistability related to the "foldover" effect. The higher the excitation frequency is, the larger the number of stable amplitudes and correlated oscillation frequencies.

c. the amplitude is mainly related to excitation frequency and the increase of the excitation amplitude weakly affects the stability of the oscillation as long as the oscillation mode is kept; the circuit compensates the excitation amplitude by changing the phase of the gated signal so that the averaged value is kept constant.

d. the time width of the gated signal should be optimized to obtain odd number of half-cycles of the excitation signal in a pulse.

e. the circuit acts correctly as a frequency divider for different shapes of the non linearity (equivalent to the variation of the shape of the single-well potential); this demonstrates the robustness of the concept.

The presented principle could firstly be used for high efficiency self-stable supplies driven with high frequencies. This application will allow class C amplifiers and consequently high efficiency of the driving; moreover higher Q-factor is observed above 10 kHz resonant frequency and higher frequency will give favorable operating conditions. In the field of instrumentation, the proposed concept can be used to stabilize the oscillation amplitude of specific sensors on the microscale (for instance, the resonating local probes in scanning near-field microscopes). In the field of memories, the amplitude could be considered as the memory level and multi-state memories become natural (the high level of integration will allow small and low cost systems). But many other applications are opened in the field of time and frequency where non linearity is daily used in both electrical and optical fields.

**Acknowledgement**
We thank Prof. Marc Gazalet for helpful documentation and discussions.